\documentstyle[twocolumn,aps,psfig,epsfig]{revtex}
\begin{document}
\draft
%\twocolumn[\hsize\textwidth\columnwidth\hsize\csname @twocolumnfalse\endcsname

\title{
Secondary Neutrinos from Tau Neutrino Interactions in Earth
}
\author{Sharada Iyer Dutta$^1$, Mary Hall Reno$^{2}$ and Ina Sarcevic$^{3}$}
\address{
$^1$Department of Physics, SUNY Stony Brook, Stony Brook, NY 
11794\\
$^2$Department of Physics and Astronomy, University of Iowa, Iowa City,
Iowa 52242\\
$^3$Department of Physics, University of Arizona, Tucson, Arizona
85721\\
}

\wideabs{
\maketitle

\begin{abstract}
\widetext
The energy dependence of ``secondary'' neutrinos from
the process $(\nu_\tau\rightarrow 
\tau\rightarrow\bar{\nu}_\mu\rightarrow \bar{\mu})$ for two input
tau neutrino fluxes ($F^0_\nu\sim E_\nu^{-1}$ and $E_\nu^{-2}$),
assumed to have been produced via neutrino oscillations
from extragalactic sources, is evaluated
to assess the impact of secondary neutrinos on upward muon rates in
a km$^3$ detector. We show that
the secondary fluxes are considerably suppressed for the steeper flux,
and even for fluxes $\sim E_\nu^{-1}$, the secondary flux will
be difficult to observe experimentally.
\end{abstract}
}
\vskip 0.1true in

\narrowtext
%\section{Introduction}

Evidence of neutrino oscillations from measurements of the atmospheric
$\nu_e$ and $\nu_\mu$ fluxes leads one to the conclusion that
$\nu_\mu\rightarrow \nu_\tau$ over distances characterized by the
radius of the Earth for neutrino energies $\sim 1$ GeV \cite{Fukuda:1998mi}.
The results of a two-flavor analysis yield $\Delta m^2=2.5\times 10^{-3}$
eV$^2$ and bi-maximal mixing, $\sin^22\theta=1$. 
Three flavor analyses are consistent with
this result \cite{Kielczewska:ia}.

Atmospheric neutrinos come from cosmic ray interactions with air nuclei,
yielding hadrons, especially mesons, which decay to neutrinos.
Neutrinos may also be produced at the sources of cosmic rays, where energetic
protons interact with nucleons and photons at the source. The pions
and kaons
produced at these distant astrophysical sources 
are parents of neutrino fluxes. Active galactic nuclei and gamma ray bursters
are two proposed astrophysical neutrino sources \cite{Learned:sw}.
Other possible sources include exotic particle annihilations \cite{crotty}.
Bi-maximal mixing, in the context of astrophysical neutrino sources,
results in flavor oscillation from the source ratio of fluxes:
$\nu_e:\nu_\mu:\nu_\tau = 1:2:0$ to flux ratios at the Earth of
$\nu_e:\nu_\mu:\nu_\tau = 1:1:1$, independent of the neutrino energy,
given that astronomical distances are so large \cite{Ahluwalia:2000fq}.

Halzen and Saltzberg pointed out in Ref. \cite{Halzen:1998be}
that high energy 
$\nu_\tau$ flux attenuation in the Earth differs from $\nu_\mu$
flux attenuation due to the fact that the $\tau$ produced in charged-current
(CC) interactions with nucleons decays before it loses energy.
For each $\nu_\tau$ lost in CC interactions, another $\nu_\tau$
appears following each $\tau$ decay, albeit at a lower energy.
Detailed evaluations of $\nu_\tau$ flux attenuation in the Earth
appear in Refs. \cite{Iyer:1999wu,Dutta:2000jv,Becattini:1999nu}.
Depending on the incident flux, the $\nu_\tau$ flux shows a degree of
``pile-up'' as neutrinos of sufficiently high energy interact in the
Earth and yield neutrinos at lower energies.

In a recent paper, Beacom, Crotty and Kolb \cite{Beacom:2001xn}
have suggested that in addition to a pile-up of tau neutrinos,
the signal of astrophysical tau neutrinos will be enhanced by
the appearance of ``secondary'' neutrinos. These secondary neutrinos
come from purely leptonic decays of $\tau$'s. The idea is that while
$\nu_\ell$ ($\ell =e,\, \mu$) fluxes starting, for example,
at nadir angle 0 are extinguished for sufficiently high energies,
they will be regenerated by the $B=0.18$ branching fraction for
$\tau\rightarrow \nu_\tau \ell {\bar {\nu}_\ell}$, the $\tau$ being
produced by $\nu_\tau\rightarrow \tau$ CC interactions.
The flux of ${\bar {\nu}_\ell}$ is less likely to be extinguished
due to the shorter path-length through the Earth (the $\nu_\tau$ already
had to travel its ``interaction distance,'') and its lower energy
due to the combined energy loss in the CC process and the decay of
the $\tau$.

The evaluations in  Ref. \cite{Beacom:2001xn} were for mono-energetic
neutrinos. In this paper, we consider a few energy dependences
for the $\nu_\tau$ fluxes. We show that the secondary fluxes are
considerably suppressed for steeply falling fluxes 
and even for
fluxes $\sim E_\nu^{-1}$, the secondary flux will be difficult to
observe experimentally.

%\section{Neutrino Attenuation}

Neutrino attenuation in the Earth is governed by a coupled 
set of partial differential equations. To illustrate, we write the coupled
equations for the $\nu_\tau$ flux.
For energy dependent
flux $F_{\nu_\tau}$ (neutrinos/(cm$^2$s\,sr\,GeV)), 
\begin{eqnarray}
&&{\partial F_{\nu_{\tau}}(E,X)\over \partial X}=
-{F_{\nu_{\tau}}(E,X)\over {\cal L}
_\nu^{int}(E)}
\\ \nonumber 
&&+\int_E^\infty dE_y \,[G^{\nu_\tau\rightarrow \nu_\tau}(E,E_y,X)
+G^{\tau\rightarrow \nu_\tau}(E,E_y,X)\,]
\\ \nonumber
&&{\partial F_{{\tau}}(E,X)\over \partial X}=
-{F_{{\tau}}(E,X)\over {\cal L}_\tau^{int}}
-{F_{{\tau}}(E,X)\over {\cal L}_\tau^{dec}}
\\ \nonumber 
&&+\int_E^\infty dE_y\,[ G^{\tau\rightarrow \tau}(E,E_y,X)
+ G^{\nu_\tau\rightarrow \tau}(E,E_y,X)\,]\ .
\end{eqnarray}
Here,
${\cal L}_\nu^{int}={1/ N_A\sigma_{\nu N}}$ \cite{gqrs98} and similarly for the
$\tau$ interaction length, and ${\cal L}_\tau^{dec}=\gamma c \tau \rho$
for density $\rho$ and Lorentz factor $\gamma=E_\tau/m_\tau$. The quantity
$X$ is the column depth (in g/cm$^2$), and for example,
\begin{equation}
 G^{\nu_\tau\rightarrow \nu_\tau}(E,E_y,X)
=\Biggl[ {F_\nu(E_y,X)\over {\cal L}_\nu^{int}}
\Biggr]{dn^{NC} \over dE}(E_y,E)\ .
\end{equation}
The cross section normalized energy distribution of
neutrinos with incident energy $E_y$ and final energy $E$ is
represented by $dn^{NC}/dE$ in Eq. (2).

There are similar equations for $\bar{\nu}_\tau$ and $\bar{\tau}$ fluxes.
In our discussion of flux ratios in the introduction, we didn't distinguish
between neutrinos and anti-neutrinos. Models predict that $F_\nu=F_{\bar{\nu}}$
to a good approximation. 
For our discussion in this section, we distinguish
between particles and antiparticles since the secondary neutrinos
are actually anti-neutrinos ($\bar{\nu}_\ell$) for incident $\nu_\tau$.
One must keep in mind that there are equal incident neutrino and antineutrino
fluxes of all three flavors in the context of neutrino oscillations.
In the event rates evaluated below, we sum both neutrino and antineutrino
contributions.

For nadir angle 0, ${\cal L}_\nu^{int}$
equals the Earth's diameter when $E_\nu\simeq 40$ TeV.
Tau neutrino pile-up and associated secondary anti-neutrino production is
relevant at this energy and higher. At very high energies, we have shown
that the effect of attenuation for a surface detector viewing at zero
nadir angle is significant even for $\nu_\tau$. For example, for
a $1/E_\nu$ flux (with a smooth cutoff at $E_\nu=10^8$ GeV), the
attenuated flux of $\nu_\tau$ at $E_\nu=10^6$ GeV is only
6\% of the incident flux at that same energy \cite{Iyer:1999wu}. 
The $\nu_e$ and $\nu_\mu$
fluxes are about 1\% of the incident flux at the same energy. 
Steeper fluxes have an even more significant attenuation and thus will have
lower relative detection rates at high energies, whether it be 
$\nu_e,\ \nu_\mu$ or $\nu_\tau$ incident fluxes.

Guided by the falling fluxes and the increased attenuation, we confine our
attention to the energy range of $E_\nu=10^3-10^8$ GeV. For
energies below $10^8$ GeV,
${\cal L}_\tau^{int}> {\cal L}_\tau^{dec}$. Starting at
$E\sim 10^8$ GeV, $G^{\tau\rightarrow
\tau}$ becomes important \cite{Dutta:2000hh}, however, we neglect it here
and confine our attention to lower energies. We also neglect the contribution
of ${\cal L}_\tau^{int}$ in Eq. (1).
With the fluxes
considered below, the error in this approximation should be small
for rates evaluated with minimum energies of $10^4-10^5$ GeV.
With these approximations,
the set of differential equations is simplified
and solved \cite{Iyer:1999wu} 
using a modification of the iterative method detailed by
Naumov and Perrone in Ref. \cite{Naumov:1998sf}.
For $\nu_\mu$ fluxes, $G^{\mu\rightarrow \mu}$ 
from electromagnetic muon energy loss effectively eliminates
any return of $\nu_\mu$ from CC interactions followed by $\mu$ decay in
the energy range of interest.

The $\tau$ flux solution to Eq. (1) is responsible for generating
the secondary neutrino flux. In the energy range of interest, one 
can write the differential equation for the $\bar{\nu}_\ell$ flux
as:
\begin{eqnarray}
&&{\partial F_{\bar{\nu}_{\ell}}(E,X)\over \partial X}=
-{F_{\bar{\nu}_{\ell}}(E,X)\over {\cal L}
_{\bar{\nu}}^{int}(E)}\\ \nonumber 
&&
%\quad 
+\int_E^\infty dE_y\,[ G^{\bar{\nu}_\ell\rightarrow \bar{\nu}_\ell}
(E,E_y,X)
+
G^{\tau\rightarrow \bar{\nu}_\ell}(E,E_y,X)\,]\ .\\
\nonumber
\end{eqnarray}

In what follows, we set $G^{\nu\rightarrow
\nu}=G^{\bar{\nu}\rightarrow\bar{\nu}}$.
At sufficiently high energies, neutrino and antineutrino interaction
rates are equal because the cross sections are dominated by the
sea quark distributions. The energy at which the interaction
length equals the column depth increases with nadir angle, so the
approximation is best at larger nadir angles.
Because the pile-up comes from higher energy
neutrino interactions, this approximation is not unreasonable, as discussed
below.

A second approximation to obtain the secondary flux is to take:
\begin{equation}
G^{\tau\rightarrow\bar{\nu}_\ell}\simeq B\cdot 
G^{\tau\rightarrow {\nu}_\tau}\ ,
\end{equation}
for $B=0.18$, the branching fraction for $\tau\rightarrow \ell$.
In fact, the $\bar{\nu}_\ell$ spectrum from $\tau$ decay is a little
softer than the $\nu_\tau$ spectrum, so this is an approximation that
will slightly overestimate the  secondary flux.

With these two approximations, one finds that the combination
$\Delta(E,X)=F_{\nu_\tau}(E,X)-F_{\bar{\nu}_\ell}(E,X)/B$
satisfies the same transport equation as $F_{\nu_\mu}(E,X)$.
At $X=0$, $\Delta(E,0)=F_{\nu_\tau}(E,0)=F_{\nu_\mu}(E,0)$,  so we
can write
\begin{equation}
F_{\bar{\nu}_\ell}(E,X)\simeq B\cdot \bigl( F_{\nu_\tau}(E,X)-F_{\nu_\mu}(E,X)
\bigr)\ .
\end{equation}
Eq. (5) will be used in what follows to approximate the secondary antineutrino
flux for two different incident spectra: $E_\nu^{-1}$ and $E_\nu^{-2}$.

In Fig. 1, we show the ratio of the attenuated flux to the incident
flux at nadir angle $\theta=0,\ 30$ and 60 degrees for 
\begin{equation}
F_\nu^0=F_\nu(E,X=0)=
N_1/E\cdot 1/(1+E/E_0)^2
\end{equation} 
where $E_0=10^8$ GeV and $N_1$ is a normalization factor, and
for $F_\nu^0\propto 1/E^2$.
The dashed curve
is the $\nu_\tau$ flux, the dotted
curve shows the attenuation
of the other two neutrino species. The solid curve, the result of 
Eq. (5), is the secondary $\bar{\nu}_\ell$ flux, were $\ell=e$ or $\mu$.
Except for the smallest nadir angles for the
$E_\nu^{-1}$ spectrum, the secondary antineutrino flux is
a small correction to the primary attenuated electron neutrino or
muon neutrino flux.
Even at nadir
angle zero, the secondary flux is negligible compared to the transmitted
primary flux for the $1/E_\nu^2$ spectrum.

One should note that even though the $\nu_\tau$ flux dominates
the primary and secondary $\nu_\mu$ fluxes, it does not dominate
the contributions to the muon event rate because of the
branching fraction $B=0.18$ of $\tau\rightarrow \mu$ together with
the effect of energy loss as the $\nu_\tau$ converts to a $\tau$ which
then decays to a $\mu$. 

%\section{Discussion}

As a quantitative illustration of the implications of the secondary
neutrino flux, we consider the muon event rate 
from the $F^0_{\nu+\bar{\nu}}
\simeq N_1/E_\nu$ case (Eq. (6)) with $N_1=10^{-13}$/(GeV
cm$^2$ s sr) for each neutrino flavor. 
The normalization factor $N_1$ is chosen
to be in line with the Waxman and Bahcall gamma ray burster flux
of Ref. \cite{Waxman:1997ti}, 
in which $N_1=4\times 10^{-13}$ (in the same units) for
the sum of all neutrino species and for $E_\nu<10^5$ GeV.
The event rates shown below are for an underground detector of 
1 km$^2$ effective area, for example, the proposed
IceCube detector \cite{icecube}. Details of the calculation 
appear in Ref. \cite{Dutta:2000jv}.
Our integrals over neutrino energies were performed 
from the minimum neutrino energy $E_\mu>10^4$ or
$10^5$ GeV, up to a
maximum neutrino energy of $10^8$ GeV. 

In Fig. 2, we include the contributions from $\nu_\mu\rightarrow \mu$
and $\nu_\tau\rightarrow \tau\rightarrow \mu$ (shown with the dashed lines)
and in addition, the corresponding antineutrino induced antimuons
from $\nu_\tau\rightarrow \tau\rightarrow\bar{\nu}_\mu\rightarrow 
\bar{\mu}$ (solid line). Anti-neutrino induced muons and antimuons are
also included. We note that 
for the $\nu_\mu
\rightarrow \mu$ rates,
doing the neutrinos and antineutrinos separately results in at most
a $\sim 10\%$
\vfil\break
 %\vfil\break

%\vfil\break
\onecolumn{
\begin{figure*}[!hbt]
\rule{0.0cm}{4.0cm}\\
\epsfxsize=17cm
\epsfbox[0 0 4096 4096]{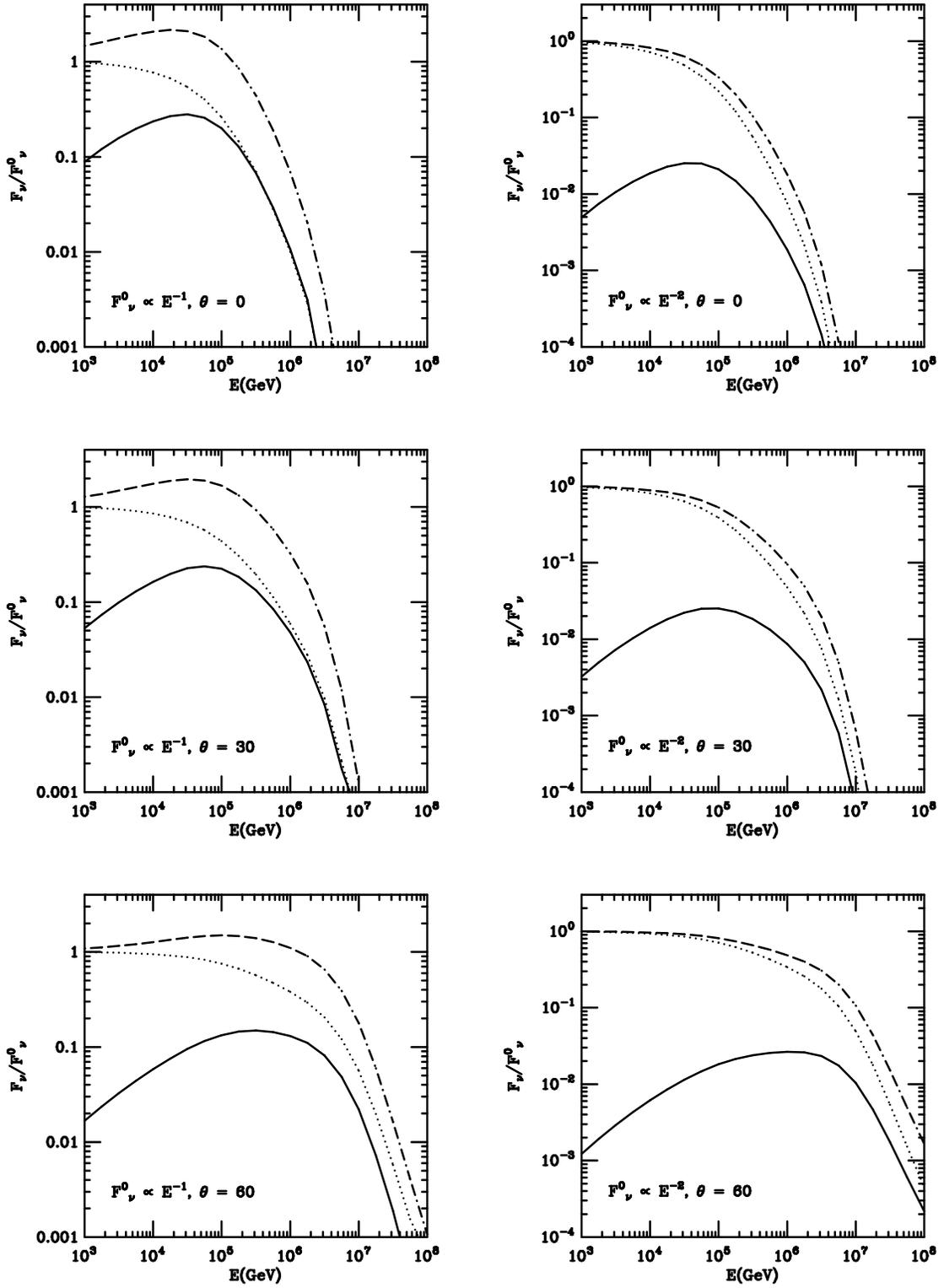}
\vspace{1cm}
\caption{The ratio of the attenuated neutrino flux for $\nu_\tau$
(dashed), $\nu_\mu=\nu_e$ (dotted) and secondary $\bar{\nu}_\ell$
(solid) to incident flux $F_\nu^0\propto E_\nu^{-1}$ (Eq. (6)) and
$F_\nu^0\propto  E_\nu^{-2}$.}
\end{figure*}
}
\twocolumn

\noindent
correction to the event rates at the energies shown here.
This small correction is an indication that our approximations to obtain
the secondary flux of neutrinos are not unreasonable.

\begin{figure}
\psfig{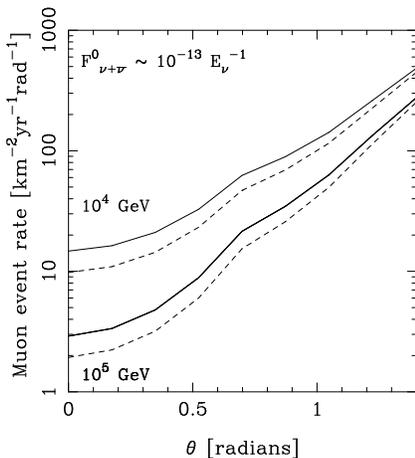}
\caption{The muon event rate for muons with energy above $E_\mu=10^4$ 
and $10^5$ GeV
originating from neutrinos with $E_\nu<10^8$ GeV from
$(\nu_\mu\rightarrow \mu)+(\nu_\tau\rightarrow \tau\rightarrow \mu)$ 
plus antiparticles (dashed)
and  additionally from secondaries via
$(\nu_\tau\rightarrow \tau\rightarrow\bar{\nu}_\mu\rightarrow \bar{\mu})$
plus antiparticles 
(solid) for the incident flux as in Eq. (6) and
$N_1=10^{-13}$/(GeV cm$^2$ s sr) for each incident neutrino
plus antineutrino
flavor.}
\end{figure}

The secondary neutrino contribution to the muon event rate has its largest
relative contribution at nadir angle zero, with an enhancement 
over the $\nu_\mu\rightarrow \mu$ plus
$\nu_\tau\rightarrow \tau\rightarrow \mu$ rate of 50-60\%
for the $1/E_\nu$ flux. At this angle, the ratio of rates of the
secondary contribution to the muon event rate relative to the event
rate of muons from tau decays is quite large, about 1.5 for 
$E_\mu>10^4$ GeV and 2.6 for $E_\mu>10^5$ GeV.
Unfortunately, this is where the event rate is smallest
and statistics are low. By a nadir angle of $\sim 60^\circ$ (1 rad), 
where the event rate is roughly a factor of 10-20 larger, depending
on the minimum muon energy, the
enhancement  in the overall muon rate is about $25\%$.
At this nadir angle, the secondary $\nu$ produced muons are equal to
the tau decay muon rate for $E_\mu>10^5$ GeV. The crossover occurs at
$\theta \sim 0.7$ rad for $E_\mu>10^4$ GeV. At the larger nadir angles,
the $\nu_\mu\rightarrow \mu$ contribution to the muon event
rate is dominant.

The normalization of the isotropic $1/E_\nu$ flux has been guided by
the Waxman-Bahcall gamma ray burster flux of Ref.
\cite{Waxman:1997ti}. 
With $E_0=10^8$ GeV in Eq. (6), our flux violates the Waxman-Bahcall
bound \cite{Waxman:1997ti} above $E_\nu\sim 10^6$ GeV, so the event
rates in Fig. 2 may be optimistic.
If the normalization $N_1$ is reasonable, then the secondary
contribution to the muon rate will be difficult to observe. Low statistics
will make it hard to have a meaningful  comparison 
between the small and large nadir angle rates. Compounding the problem
is that one does not expect to know the input flux energy dependence
or normalization exactly.

For more rapidly falling fluxes, the contribution
of secondary neutrinos to the event rates is smaller. The attenuated
fluxes shown
in Fig. 1 for $1/E_\nu^2$ yield secondary enhancements which are quite
small. The secondary muon rate is only about 10-15\%  of the primary
$\nu_\mu\rightarrow \mu$ rate at nadir angle zero. 
Typical theoretical neutrino fluxes have spectra that lie somewhere between
the $1/E_\nu$ and $1/E_\nu^2$ cases in this energy range \cite{Learned:sw}.

In summary, the energy dependence of the incident tau neutrino flux
is crucial in evaluations of the implications of $\nu_\tau$ interactions
to regenerate $\nu_\tau$ and secondary $\bar{\nu}_\mu$ and $\bar{\nu}_e$.
Neutrino fluxes are attenuated due to their passage through the Earth,
even the tau neutrinos, thus moderating tau neutrino contributions
to secondary neutrino fluxes.
Our evaluation has relied on approximations
including Eq. (4), setting ${\cal L}_\tau^{int}=0$ and $G^{\nu\rightarrow
\nu}=G^{\bar{\nu}\rightarrow \bar{\nu}}$. If muon rates are determined to
be large, even near nadir angle zero, then a more detailed evaluation
of the secondary flux may be in order.
With our current theoretical expectations for flux normalizations,
however, the secondary neutrinos coming from $\tau$
decays will be difficult to observe experimentally, as they contribute
significantly to a muon excess only at small
nadir angles where the fluxes are already strongly attenuated and
for spectra like $1/E_\nu$.

%\vskip 0.1true in

%\leftline{\bf Acknowledgments}  

%\vskip 0.1true in

We thank J. Beacom for discussions.
The work of S.I.D. has been supported in part by NSF Grant
0070998. The work of I.S. has been supported in part by the DOE under
contracts DE-FG02-95ER40906 and DE-FG03-93ER40792.  The work of
M.H.R. has been supported in part by NSF Grant
No.  PHY-9802403 and DOE under contract
FG02-91ER40664.  M.H.R. thanks the Fermilab Theory Group
for its hospitality.

\end{document}